\documentclass{ws-procs9x6}

\begin{document}
\title{Point-Form Approach to Baryon Structure}
\author{W. PLESSAS
%\footnote{}
}
\address{Theoretische Physik, Institut f\"ur Physik \\
Universit\"at Graz, \\ 
Universit\"atsplatz 5,
A-8010 Graz, Austria\\ 
E-mail: plessas@uni-graz.at}
\maketitle
\abstracts{
A critical discussion is given of the results for baryon electromagnetic
and axial form factors
obtained from relativistic constituent quark models in the framework
of Poincar\'e-invariant quantum mechanics. The primary emphasis lies on
the point-form approach. First we summarize the predictions of the 
Goldstone-boson-exchange constituent quark model for the electroweak 
nucleon structure when using a spectator-model current in point form.
Then the influences of different dynamics inherent in various kinds of 
constituent quark models (Goldstone-boson-exchange, 
one-gluon-exchange, instanton-induced interactions) are
discussed. Finally the point-form results
are compared to analogous predictions calculated in instant form. 
Relativistic effects are always of sizeable magnitude. A 
nonrelativistic approach is ruled out. The instant-form results are 
afflicted with severe shortcomings. In the spectator-model 
approximation for the current, only the point-form results appear to
be reasonable a-priori. In fact, the corresponding quark model predictions
provide a surprisingly good description of all elastic electroweak
observables in close agreement with existing experimental data, 
specifically for the Goldstone-boson-exchange constituent quark model.   
}
\section{Introduction}
Constituent quark models (CQMs) have become a reliable concept for 
the description of hadron spectroscopy. Specifically the low-lying 
spectra of the light and strange baryons have experienced a reasonable 
explanation by respecting the spontaneous breaking of chiral 
symmetry of quantum chromodynamics (QCD) in the dynamics employed for 
the effective interaction between constituent quarks. In this respect, 
the so-called Goldstone-boson-exchange (GBE) CQM\cite{Glozman:1998ag}
has proven especially adequate\cite{Glozman:1998fs}. Consequently it 
appears essential to include the relevant symmetries of low-energy QCD 
in the construction of any CQM. It is of similar importance to observe 
the symmetry requirements of special relativity. In order to obtain 
relativistic predictions for observables, any CQM must be based on a 
dynamical concept (e.g., a relativistic mass operator or an equivalent
Hamiltonian) invariant under the 
transformations of the Poincar\'e group.    

In following a relativistic quantum-mechanical treatment of few-quark 
systems one must make a choice of the formalism to be applied. The 
different approaches are distinguished by the specific stability 
subgroups of the Poincar\'e group in case of an interacting
system\cite{Dirac:1949,Keister:1991sb}. 
The point form is characterized by four generators dependent on 
interactions, namely, the components of the 
four-momentum. The stability subgroup of the instant form has the 
same dimension (with the Hamiltonian and the three generators of the 
Lorents boosts dependent on interactions). In case of the front form, 
only three generators are interaction-dependent. 

Until a few years ago the point form had been the approach least
frequently followed, even though it has specific advantages. For 
instance, one can easily and accurately apply Lorentz boosts, since
their generators remain purely kinematical. Following the works by
Klink et al.\cite{Klinketal},
the Graz-Pavia collaboration has applied the point form to the calculation
of electromagnetic and axial form factors of the
nucleon\cite{Wagenbrunn:2000es,Glozman:2001zc,Boffi:2001zb}. One has 
obtained very remarkable results. The direct predictions of the GBE CQM, 
calculated with the nucleon wave functions just as obtained from the
quark model, have been found in close agreement with experimental data
in all instances. The behaviour of these results is therefore rather 
distinct from corresponding results obtained before in other 
approaches such as the front form (see, for example, refs.\cite{Rome}).
There, one needed quark form factors in order to bring 
the theoretical predictions into the vicinity of the experimental data.

Below I first summarize the characteristics of the point-form results for 
the electroweak structure of the nucleons.
Then I compare the covariant results with 
the nonrelativistic ones, consider different CQMs (wave functions),
and contrast the point form to the instant form. In the discussion a
few observations are made also with regard to relativistic invariance
(frame independence) and current conservation.

\section{Point-Form Results}

\begin{figure}
\begin{center}
\includegraphics[width=0.75\hsize,bb=35 14 339 253,clip=]{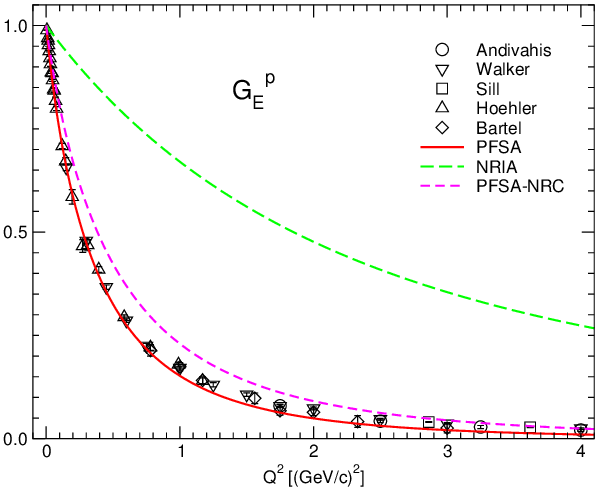}
\includegraphics[width=0.75\hsize,bb=35 14 339 253,clip=]{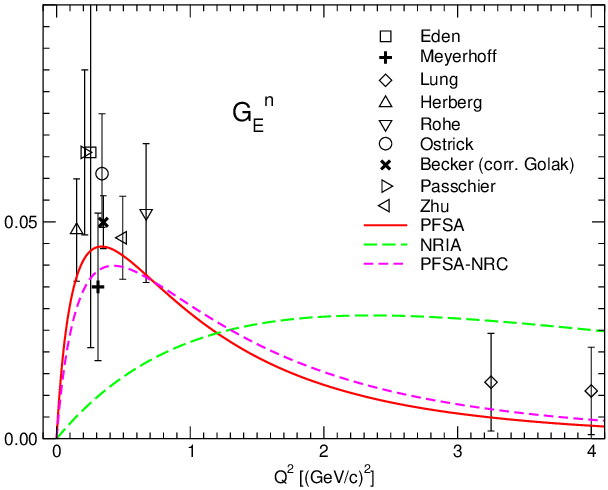}
\caption{Proton and neutron electric form factors as predicted by the 
GBE CQM\protect\cite{Glozman:1998ag}. \label{elgbe}}
\end{center}
%\vspace*{-0.2cm}
\end{figure}

\begin{figure}[t]
\begin{center}
\includegraphics[width=0.75\hsize,bb=35 14 339 253,clip=]{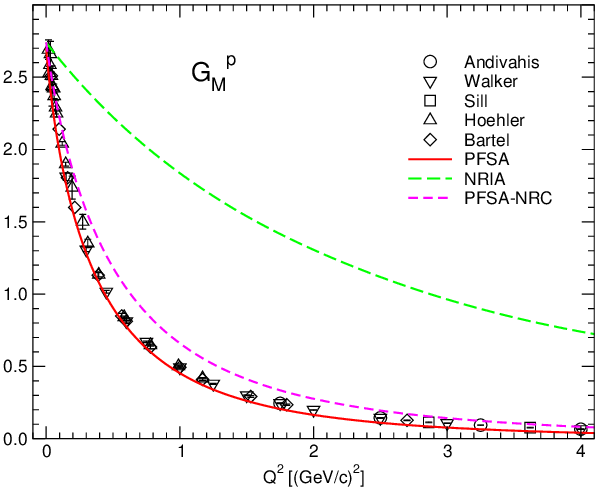}
\includegraphics[width=0.75\hsize,bb=35 14 339 253,clip=]{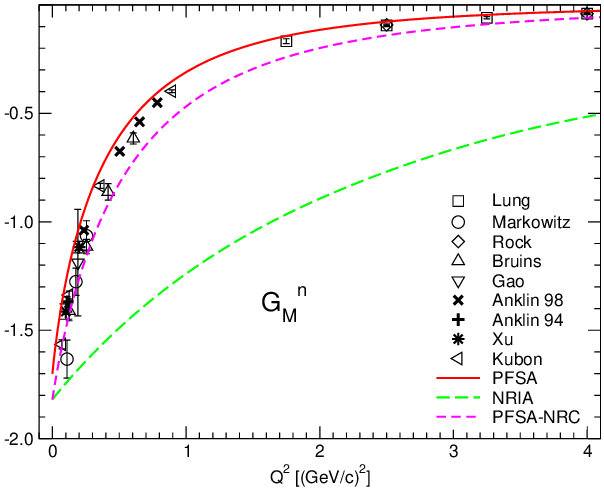}
\caption{Proton and neutron magnetic form factors as predicted by the 
GBE CQM\protect\cite{Glozman:1998ag}. \label{maggbe}}
\end{center}
%\vspace*{-0.2cm}
\end{figure}

\begin{figure}[t]
\begin{center}
\includegraphics[width=0.75\hsize,bb=35 14 339 253,clip=]{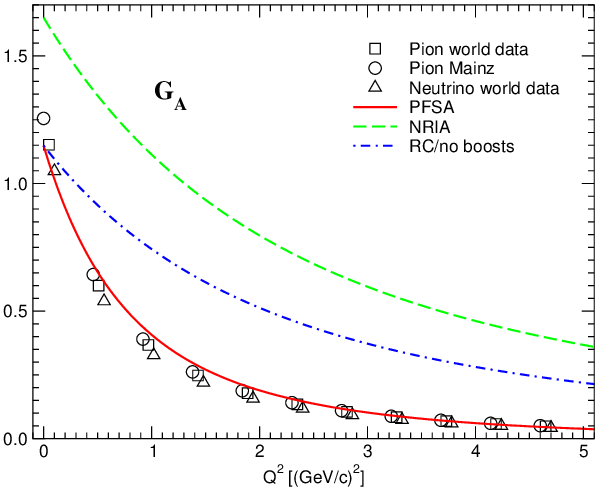}
\includegraphics[width=0.75\hsize,bb=35 14 339 253,clip=]{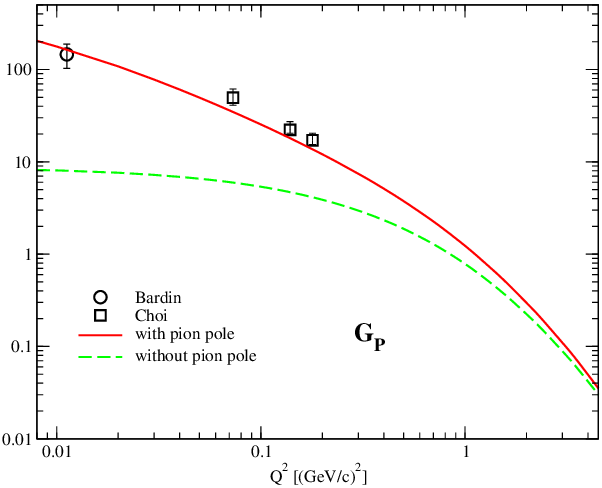}
\caption{Nucleon axial and induced pseudoscalar form factors as
predicted by the GBE CQM\protect\cite{Glozman:1998ag}. \label{axgbe}}
\end{center}
\vspace*{-0.2cm}
\end{figure}

Let us first have a look at the predictions of the GBE
CQM\cite{Glozman:1998ag} for the nucleon electromagnetic form factors in 
figs. \ref{elgbe} and \ref{maggbe} and for the axial as well as induced 
pseudoscalar form factors in fig. \ref{axgbe}. There the covariant results
obtained in point-form spectator approximation (PFSA) are displayed.
The direct predictions of the GBE CQM are
immediately found in reasonable agreement with the available
experimental data up to momentum transfers of $Q^{2} \sim 4$ 
GeV$^{2}$. On the other hand, the results calculated in nonrelativistic
impulse approximation (NRIA) fall short in every respect. In order 
to demonstrate the boost effects we also show the results that come 
out if one uses a nonrelativistic current but includes the 
boosts according to the point form (PFSA-NRC). For the axial 
form factor, instead, we give the results for the case when a relativistic 
current is employed but no boosts are included (RC/no boosts). With 
regard to the induced pseudoscalar form factor a comparison is given 
to the case when the pion pole is neglected; evidently, one then 
misses contributions of more than an order of magnitude. From all of 
these results one learns that relativity is of utmost importance and 
the pion degrees of freedom play an essential role.

\begin{figure}[t]
\begin{center}
\includegraphics[width=0.75\hsize,bb=35 14 339 253,clip=]{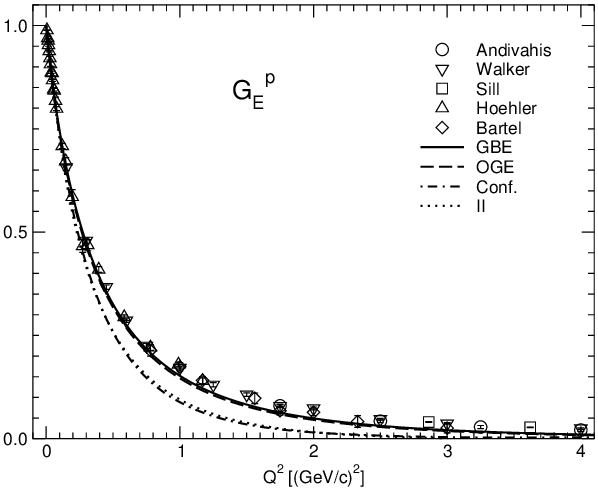}
\includegraphics[width=0.75\hsize,bb=35 14 339 253,clip=]{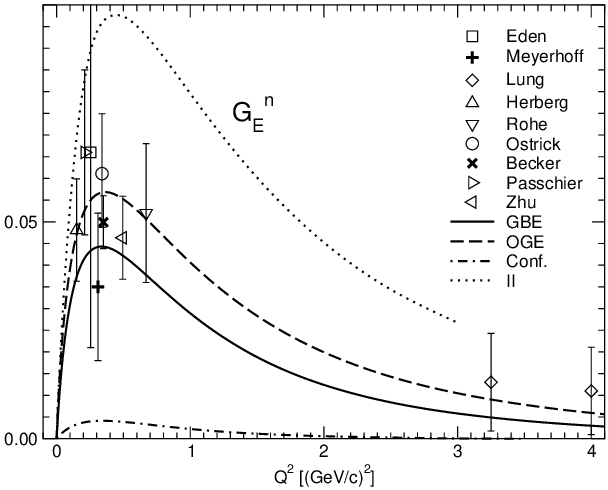}
\caption{Comparison of proton and neutron electric form factors as
predicted by the GBE\protect\cite{Glozman:1998ag},
OGE\protect\cite{Theussl:2000sj}, and II\protect\cite{Loring:2001kx} CQMs
and the case with the confinement potential only. \label{elcomp}}
\end{center}
%\vspace*{-0.2cm}
\end{figure}

\begin{figure}[t]
\begin{center}
\includegraphics[width=0.75\hsize,bb=35 14 339 253,clip=]{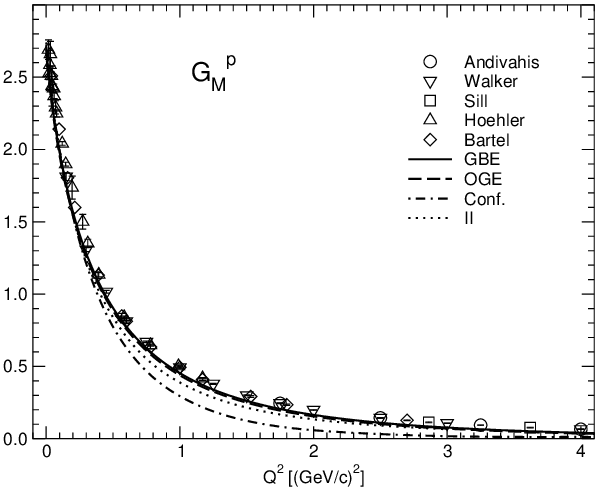}
\includegraphics[width=0.75\hsize,bb=35 14 339 253,clip=]{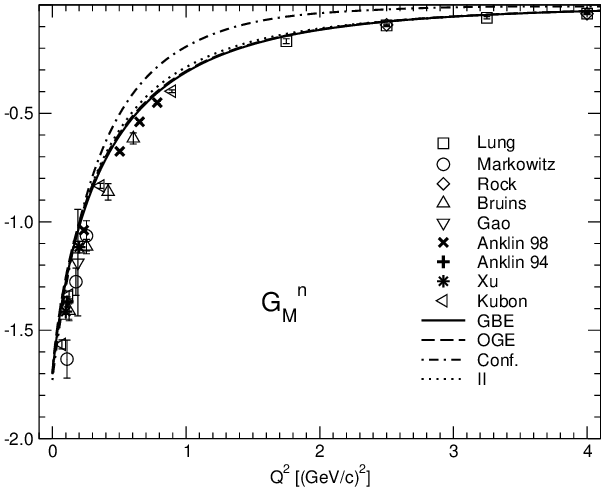}
\caption{Comparison of proton and neutron magnetic form factors as
predicted by the GBE\protect\cite{Glozman:1998ag},
OGE\protect\cite{Theussl:2000sj}, and II\protect\cite{Loring:2001kx} CQMs
and the case with the confinement potential only. \label{magcomp}}
\end{center}
%\vspace*{-0.2cm}
\end{figure}

How important are the specific dynamics prevailing in a certain CQM? In 
figs. \ref{elcomp} and \ref{magcomp} we give a comparison of the PFSA 
predictions of the GBE CQM\cite{Glozman:1998ag}, of the 
one-gluon-exchange (OGE) CQM after Bhaduri-Cohler-Nogami\cite{Bhaduri:1981pn}
in the relativistic parametrization by Theu{\ss}l et 
al.\cite{Theussl:2000sj}, and of the instanton-induced (II) CQM by the
Bonn group\cite{Loring:2001kx}, which is treated in a Bethe-Salpeter 
approach; in addition the case with the 
confinement interaction only is shown. One sees that the dynamical 
influences are rather weak once a realistic nucleon wave function is 
produced. In particular, the kind of hyperfine interaction (GBE or OGE
or II) is not so decisive, at least not for the nucleon ground state. 
If only the confinement interaction is present, however, one faces 
severe shortcomings especially with respect to the neutron form factors.
Above all the neutron electric form factor (fig. \ref{elcomp}) is
dependent on a small 
mixed-symmetry spatial component in the wave function. If it is 
absent, like in the case with the confinement interaction only, one 
practically gets a zero result. In this context we have not shown a 
comparison for the induced pseudoscalar form factor. As explained 
above it requires the pion-pole contributions, which cannot 
consistently be implemented neither for the OGE nor the II CQMs.

\begin{figure}[t]
\begin{center}
\includegraphics[width=0.7\hsize,bb=35 14 339 253,clip=]{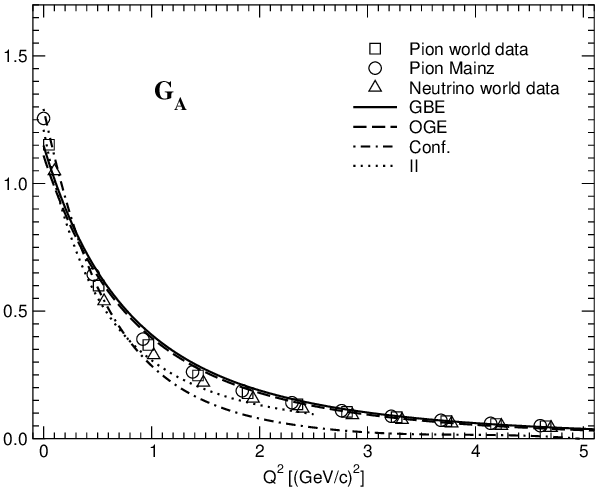}
\caption{Comparison of axial form factors as
predicted by the GBE\protect\cite{Glozman:1998ag},
OGE\protect\cite{Theussl:2000sj}, and II\protect\cite{Loring:2001kx} CQMs
and the case with the confinement potential only. \label{axcomp}}
\end{center}
%\vspace*{-0.2cm}
\end{figure}

We have not addressed the electric radii and the magnetic moments 
here. They follow from the electric and magnetic form factors in the 
limit $Q^{2} \to 0$. The corresponding results have already been
calculated not only for the nucleons but also for all other octet and 
decuplet baryon ground states\cite{BWP:2004}. Again the direct 
predictions (of the GBE CQM) in PFSA are immediately found to be 
reasonable and in good agreement with experiment in all cases
whenever data exist. Relativistic effects are of considerable 
importance also for the electric radii and magnetic moments. This may 
appear strange at first sight, since we deal here with observables 
in the limit of zero momentum transfer. Nevertheless boost effects 
bring about sizeable contributions, and a nonrelativistic theory is 
bound to fail even for these quantities\cite{BWP:2004}.

\section{Comparison of Point-Form and Instant-Form Results}

\begin{figure}[t]
\begin{center}
\includegraphics[width=0.7\hsize,bb=45 42 660 507,clip=]{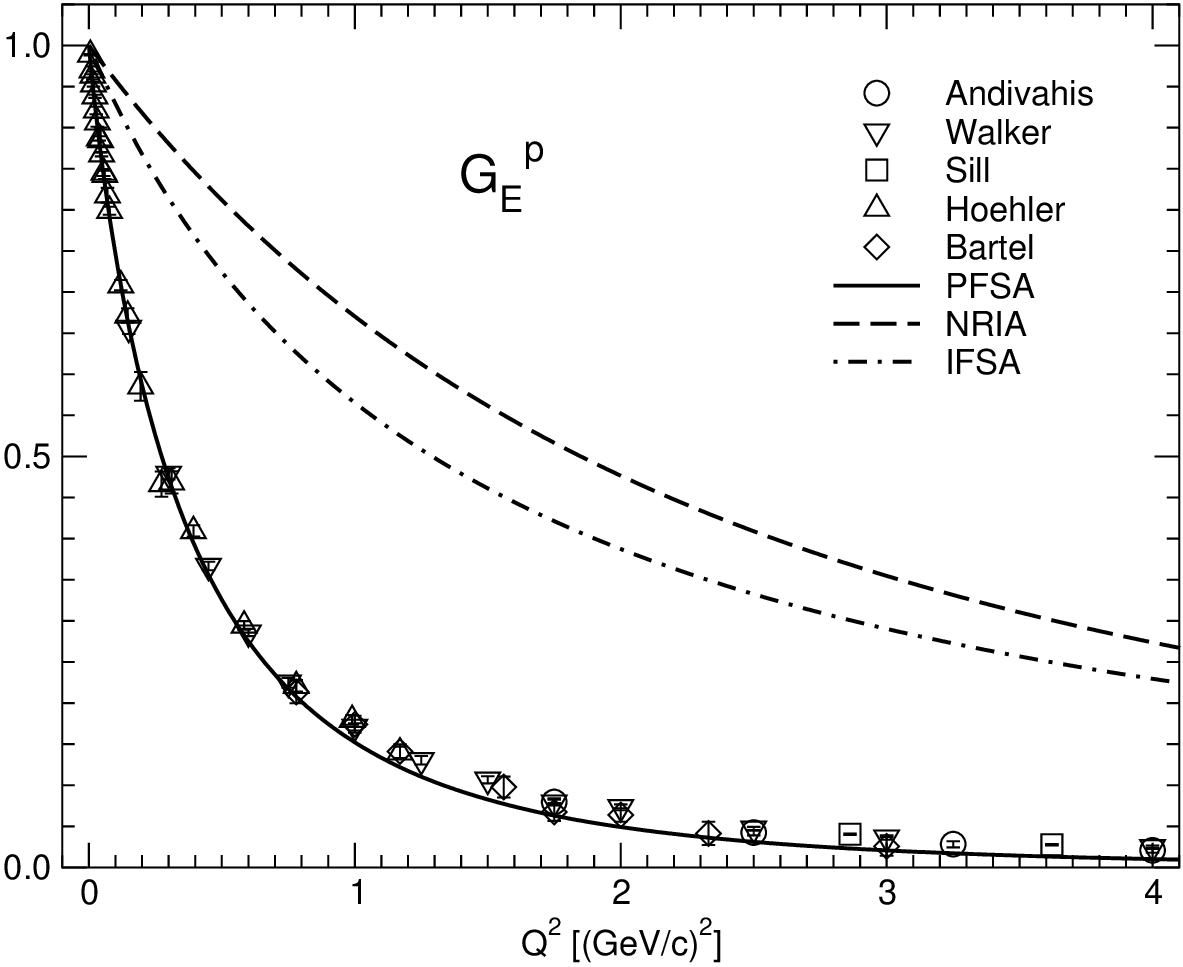}
\includegraphics[width=0.7\hsize,bb=45 42 660 507,clip=]{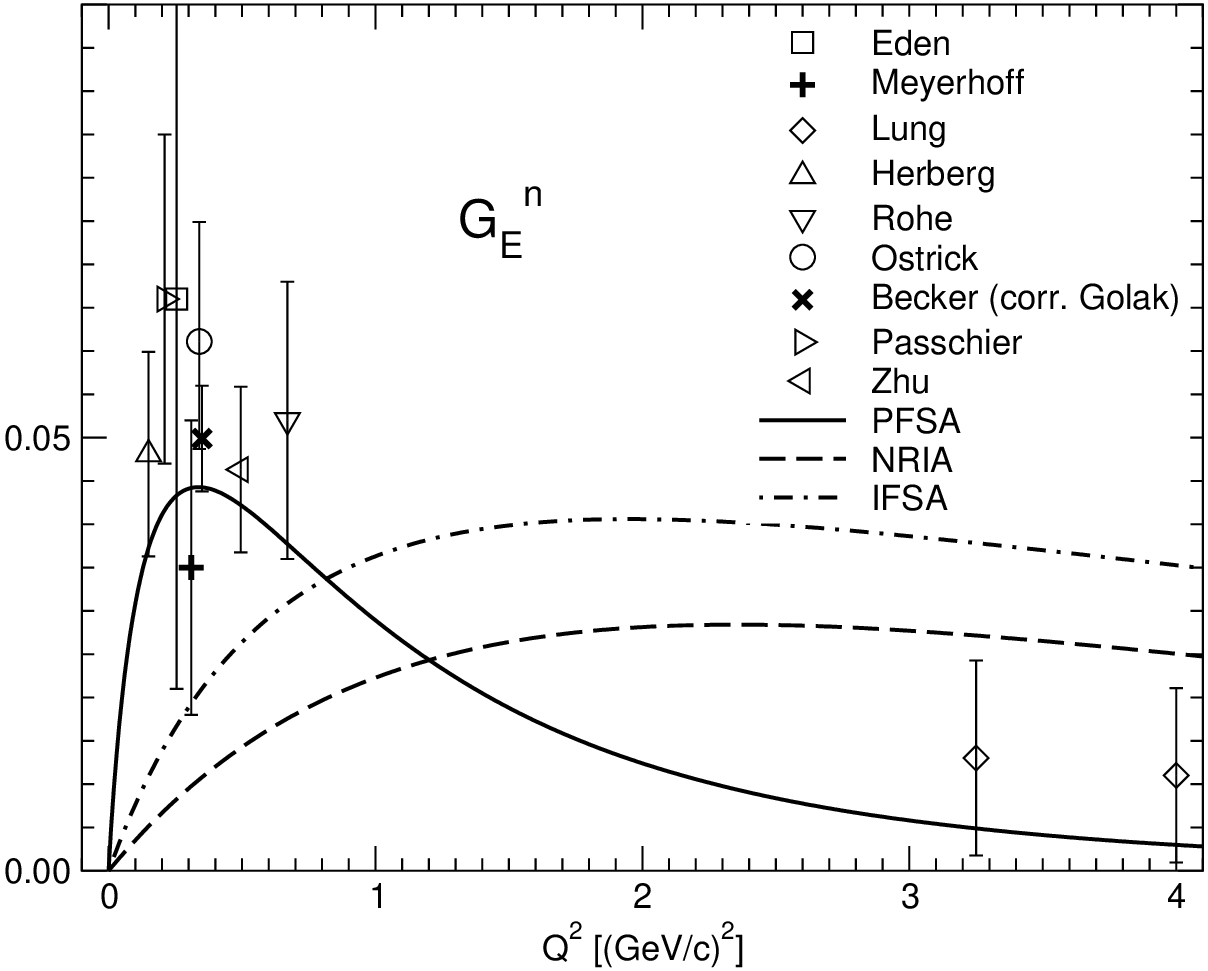}
\caption{Comparison of proton and neutron electric form factors of 
the GBE CQM calculated in PFSA and IFSA as well as in NRIA.
\label{elinst}}
\end{center}
%\vspace*{-0.2cm}
\end{figure}

\begin{figure}
\begin{center}
\includegraphics[width=0.7\hsize,bb=45 42 660 507,clip=]{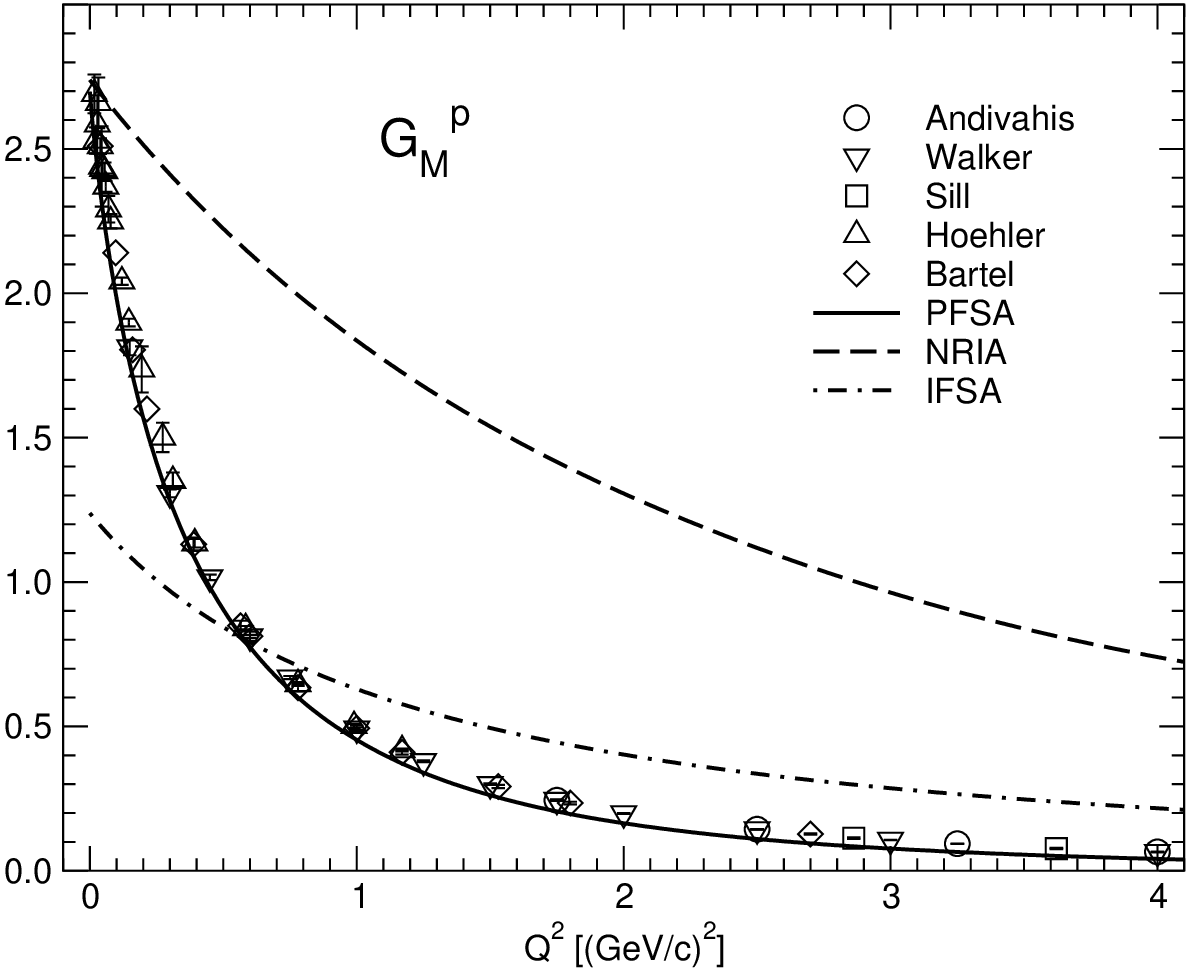}
\includegraphics[width=0.7\hsize,bb=45 42 660 507,clip=]{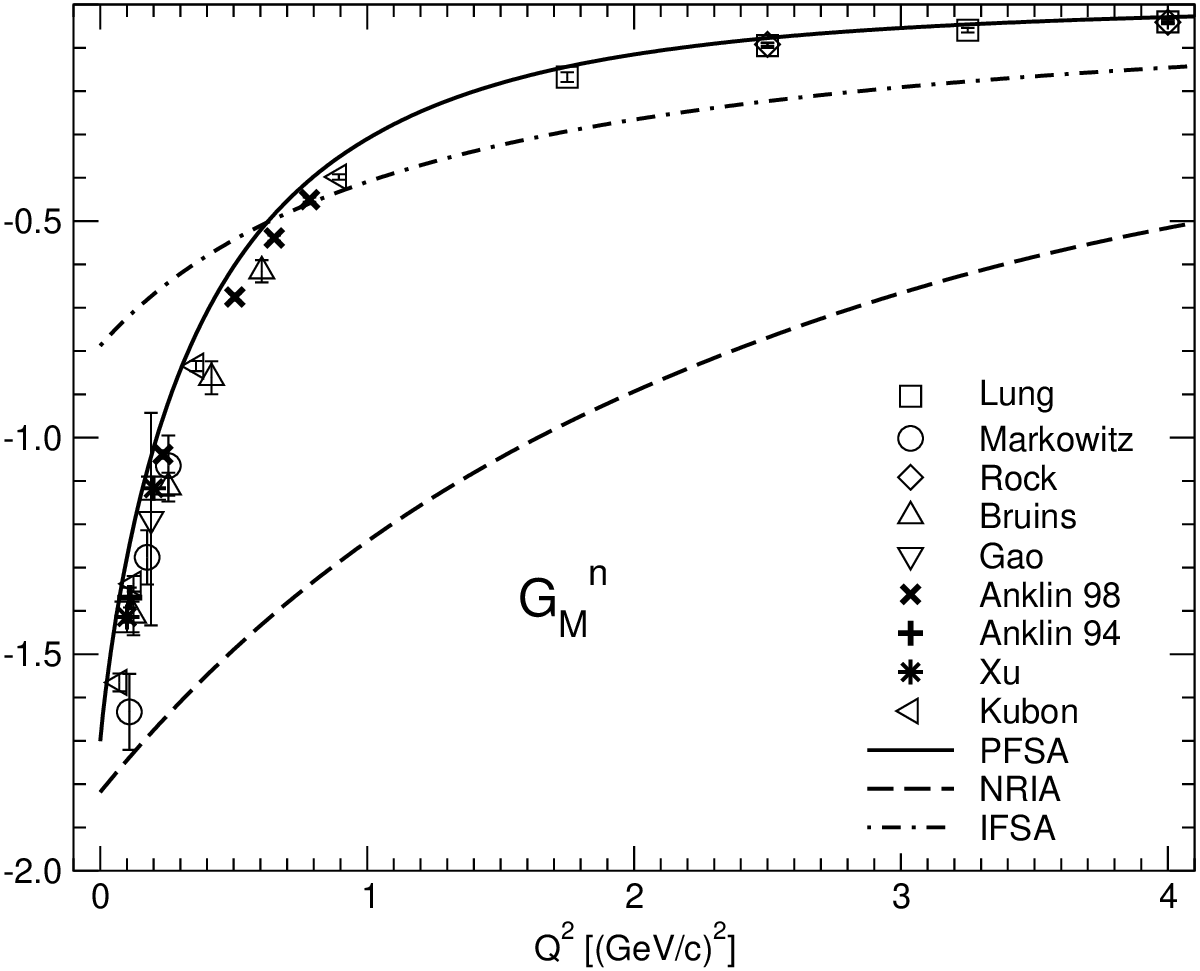}
\caption{Comparison of proton and neutron magnetic form factors of 
the GBE CQM calculated in PFSA and IFSA as well as in NRIA.
\label{maginst}}
\end{center}
%\vspace*{-0.2cm}
\end{figure}

In view of the solid performance of the relativistic approach along 
the point form one has to ask why these surprising results come out 
(whenever a realistic wave function is employed). One has to bear in 
mind that the theory is by no means complete, since only a model 
current is used, namely, the so-called PFSA current. Of course, this 
model current is certainly not a one-body current but still
the corresponding calculation may lack 
sizeable contributions from further types of few-body currents. In 
order to elucidate the properties of the point form in the spectator 
model, we have performed a completely analogous study in instant form. 
In figs. \ref{elinst} and \ref{maginst} we present a comparison of the 
results obtained with the GBE CQM in PFSA and in instant-form spectator
approximation (IFSA); in addition the NRIA (from figs. \ref{elgbe} and 
\ref{maggbe}) is repeated. It is seen that the IFSA results remain far 
away from a reasonable description of the nucleon electromagnetic
form factors. In fact, the IFSA results fall closer to the NRIA than 
to the PFSA (and thus to the experimental data), especially for the 
electric form factors. While this comparison is given for the 
Breit-frame calculations, one has to note that the instant-form 
results in the spectator-model approximation are
frame-dependent. This makes them particularly 
questionable. A serious requirement of a relativistic theory is thus 
violated. In contrast, the point-form results are frame-independent. 
They are manifestly covariant even in the spectator-model
approximation for the current.  

Another criterion for a reliable theoretical approach to 
electromagnetic form factors is current conservation. We have checked 
the fulfillment of the continuity equation in case of the PFSA. In the 
range of momentum transfers considered here, the violation of current 
conservation remains below 1~\%! This is a satisfying observation 
though it does not definitely tell that two- and three-body 
currents would ultimately be small.

\section{Conclusions}

From the present studies one can learn several important lessons. First of 
all it is evident that a nonrelativistic CQM is by no means adequate 
to describe the properties of hadrons, not even in the domain of low 
energies or momentum transfers. Second, an approach following 
relativistic (Poincar\'e-invariant) quantum mechanics turns out to be
justified and convenient. It allows to implement the symmetry
requirements of special 
relativity and is not confronted with the problems of a 
field-theoretic approach (such as truncations of infinite series, 
discretizations of integrations, etc.). Specifically the point-form 
approach seems to bring about a number of advantages. It guarantees 
a-priori for covariance, allows to solve the dynamical equations 
rigorously, and keeps the violation of current conservation very small;
in practice, it is negligible in the domain of momentum
transfers considered here. 
The IFSA, on the other hand, is affected by severe theoretical 
shortcomings. Most embarrassing is the frame dependence. It makes the 
instant-form approach in the spectator approximation very questionable
if not completely inadequate.

Certainly, at this instance, we are also left with a number of open 
problems. Even though the PFSA results provide a consistent description 
of all aspects of the electroweak structure of the light and strange 
baryon ground states, one must not forget that the approach relies on 
simplifying assumptions and is by no means complete. It is also 
distinct from a field-theoretic treatment. Obviously, one may ask for 
the contributions of two- and three-body currents. The approximate 
fulfillment of current conservation in the PFSA may be taken as a
hint that these contributions might indeed be small. However, this must
still be proven by performing calculations with a more elaborate
or even the complete current operator. Until this 
problem is settled one can also not definitely conclude on a possible 
structure of constituent quarks and/or a finite extension of
the interaction vertices. It is also clear that due to their unitary 
equivalence\cite{SS:1978} all forms of relativistic quantum mechanics must 
lead to the same results once a full calculation is performed. The 
contributions missing beyond the present spectator-model calculations 
should then turn out of different magnitudes in the point and instant
forms (and, of course, also in the front form).

Beyond the elastic form factors a number of further observables 
remain to be studied. The framework of Poincar\'e-invariant quantum 
mechanics is also applicable to inelastic processes such as 
transition form factors etc. Further important insights in the 
performance of relativistic CQMs and the adequacy of the relativistic 
quantum-mechanical approach may thus be obtained.

\vspace{-3mm}
\section*{Acknowledgment}

\noindent{The results discussed in this paper rely on essential 
contributions by my colleagues W. Klink (Iowa), S. Boffi and M. 
Radici (Pavia), as well as K. Berger, L. Glozman and especially
R. Wagenbrunn (Graz). This work was supported by the Austrian Science 
Fund (Projects P14806 and P16945).}

\end{document}